\documentstyle[11pt, appb,axodraw,epsfig,epsf]{article}
\begin{document}

\newcommand{\symbolfootnote}{\renewcommand{\thefootnote}
        {\fnsymbol{footnote}}}
\renewcommand{\thefootnote}{\fnsymbol{footnote}}
\newcommand{\alphfootnote}
        {\setcounter{footnote}{0}
         \renewcommand{\thefootnote}{\sevenrm\alph{footnote}}}
\newcommand{\ba}{\begin{eqnarray}}
\newcommand{\ea}{\end{eqnarray}}
\newcommand{\crn}{\nonumber \\}
\newcommand{\gm}{\gamma}
\newcommand{\Gm}{\Gamma}
\newcommand{\e}{\varepsilon}
\newcommand{\trg}[3]{
\begin{picture}(35,30)(5,13)
\Line(5,5)(20,30)
\Line(35,5)(20,30)
\Line(5,5)(35,5)
\Line(5,5)(3,3)
\Line(35,5)(38,3)
\Line(20,30)(20,32)
\Text(20,0)[]{$\scriptstyle #1$}
\Text(13,25)[r]{$\scriptstyle #2$}
\Text(27,25)[l]{$\scriptstyle #3$}
\end{picture}}
\newcommand{\trgleft}[3]{
\begin{picture}(35,30)(5,13)
\Line(5,5)(20,30)
\Line(35,5)(20,30)
\Line(5,5)(35,5)
\DashLine(5,5)(-10,5){2}
\Line(-10,5)(-13,3)
\Line(35,5)(38,3)
\Line(20,30)(20,32)
\Text(-3,0)[]{$\scriptstyle -1$}
\Text(20,0)[]{$\scriptstyle #1$}
\Text(13,25)[r]{$\scriptstyle #2$}
\Text(27,25)[l]{$\scriptstyle #3$}
\end{picture}
                        }

\centerline{ MULTI-LOOP CALCULATIONS IN THE STANDARD MODEL:}
\centerline{ techniques and applications
\footnote{ Presented by J. Fleischer at 21th School of
Theoretical Physics, Ustro\'n, Poland, September 1997.}.}
\baselineskip=22pt
\centerline{ J.~FLEISCHER\footnote{~E-mail:
fleischer@physik.uni-bielefeld.de}, M.~TENTYUKOV\footnote
{
On leave of absence from Joint Institute for Nuclear Research,
141980 Dubna, Moscow Region, Russian Federation.~E-mail:
tentukov@thsun1.jinr.dubna.su}%
\addtocounter{footnote}{2}%
\footnotemark[\value{footnote}]%
\addtocounter{footnote}{-2}%
~~and~~O.~L.~VERETIN\footnote{~E-mail:veretin@physik.uni-bielefeld.de
}\footnote{
Supported by Bundesministerium f\"ur Forschung und Technologie
under PH/05-7BI92P 9.
}}

\baselineskip=13pt
\centerline{Fakult\"at f\"ur Physik, Universit\"at Bielefeld}
\baselineskip=12pt
\centerline{D-33615 Bielefeld, Germany}
\vspace{0.9cm}

{\bf Abstract:} We present a review of the Bielefeld-Dubna activities on
the multiloop calculations. In the first part a C-program
DIANA (DIagram ANAlyser) for the automation of Feynman diagram 
evaluations is presented, in the second part various techniques for the
evaluation of scalar diagrams are described, based on the Taylor
expansion method and large mass expansion.

\section{Automation of  Feynman diagram evaluation}
   Recent high precision experiments require, on the side of the theory,
high-precision calculations resulting in the evaluation of higher
loop dia\-grams in the Standart Model (SM).  
For specific processes thousands of multiloop Feynman
dia\-grams do contribute, and it turns out to be impossible to perform 
these calculations by hand. This makes the request for automation a
high-priority task. 

Several different packages
have been developed with different areas of applicability.
For example,
FEYNARTS / FEYNCALC \cite{FeynmArts} are MATHEMA\-TICA packages convenient
for various aspects of the calculation of radiative corrections in the SM.
There are several FORM packages for evaluating multiloop diagrams, like
MINCER \cite{MINCER}, and a package \cite{leo96} 
for the calculation of 3-loop bubble integrals with a mass.
Other automatic packages are
GRACE \cite{GRACE} and COMHPEP \cite{CompHep},
which partially perform full calculations, from the process definition 
to the cross-section values. 

A somewhat different approach is persued  by
XLOOPS \cite{XLoops}. 
A graphical user interface
makes XLOOPS an `easy-to-handle' program package, but is mainly aimed 
to the evaluation of single diagrams.
To deal with thousands of diagrams, it is neccessary to use 
special techniques like databases and special controlling programs.
In \cite{Vermaseren} for evalua\-ting more than 11000 diagrams
the special database-like program MINOS was developed.
It calls the relevant FORM programs, waits until they fi\-nished, picks 
up their results and repeats the process without any human interference.

All these packages have different efficiency in different domains.
It seems impossible to develop an universal package, which 
will be effective for all tasks.
This point of view motivated us to seek our own
way of automatic evaluation of Feynman diagrams. 

Our first step is dedicated to the automation of 
the muons two-loop anomalous magnetic moment (AMM) 
${\frac{1}{2}(g-2)}_{\mu}$. For this purpose
the package TLAMM was developed \cite{TLAMM}.
The algorithm is implemented
as a FORM-based program package. For generating and automatically
evaluating any number of two-loop self-energy diagrams, a special
C-program has been written. This program creates the initial
FORM-expression for every diagram generated by 
QGRAF \cite{QGRAF}, executes the
corresponding subroutines and sums up the various contributions.
In the SM 1832 two-loop diagrams contribute in this case. 
The calculation of the bare diagrams is finished.
For the purpose of demonstration, we
have applied TLAMM to a closed subclass of diagrams of the SM
which we refer to as ``toy'' model. Some details of the calculation
are presented for this case.

 A more general project called DIANA (DIagram ANAlyser) \cite{FT} for the
evalua\-tion of Feynman diagrams is being finished by our group at
present and will also be shortly described below.

\subsection{The toy model}

We considered as a toy example the model involving a light charged spinor 
$\Psi$, a  photon $A_\mu$, and a heavy neutral scalar field $\Phi$.  
The scalar has triple $\left( g \right)$ and quartic $\left( \lambda \right)$
self-interactions, and the Yukawa coupling $\left( y \right)$ to the
spinor.  The Lagrangian of this model reads (in Euclidean
space-time)

\begin{eqnarray}
{\it L} & = & \frac{1}{2} \partial_\mu \Phi \partial^\mu \Phi
+ \frac{1}{2} M^2 \Phi^2 - \frac{g}{3!} \Phi^3
- \frac{\lambda}{4!} \Phi^4
+ \frac{1}{4} \left(\partial_\mu A_\nu - \partial_\nu A_\mu \right)^2
\nonumber \\ &&
+ \frac{1}{2 \alpha} \left( \partial_\mu A^\mu \right)^2
+ \bar{\Psi} \left( \hat{\partial} + i e \hat A+ m \right) \Psi
- y \Phi \bar{\Psi} \Psi
\label{toy-model}
\end{eqnarray}

\noindent
where $e$ is the electric charge and $\alpha$ is a gauge fixing parameter.

\begin{figure}[h]
\centerline{\vbox{\epsfysize=100mm \epsfbox{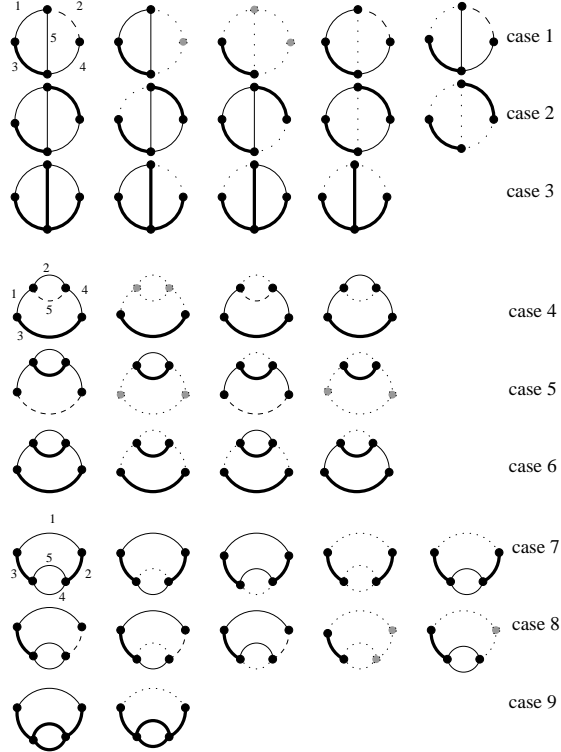}}}
\caption{\label{fig33} The prototypes and their subgraphs
contributing to the large mass expansion in the toy model.
Thick, thin and dashed lines correspond to
the heavy-mass, light-mass, and massless propagators, respectively.
Dotted lines indicate the lines omitted in the subgraph $\gamma$.}
\end{figure}
  There are $40$ diagrams contributing to the two-loop AMM of the fermion.  
The calculation of the AMM reduces, after differentiation
and contractions with projection operators, to diagrams of
the propagator type with external momentum on the fermion mass shell
(for details see \cite{projector}). After performing the Dirac 
and Lorentz algebra, all diagrams can be reduced to some set of
master integrals. 

  For the evaluation of these integrals we use the asymptotic 
expansion in large masses \cite{asymptotic}. For a given scalar
graph $G$ the expansion in large mass is given by the formula
\begin{equation}
F_G(q, M ,m, \varepsilon) \stackrel{M \to \infty}{\sim }
\sum_{\gamma} F_{G/\gamma}(q,m,\varepsilon) \circ
T_{q^{\gamma}, m^{\gamma}}
F_{\gamma}(q^{\gamma}, M ,m^{\gamma}, \varepsilon),
\label{Lama}
\end{equation}
\noindent
where $\gamma$'s are subgraphs involved in
the asymptotic expansion, $G/\gamma$ denotes shrinking of $\gamma$ to a
point; $F_{\gamma}$ is the Feynman integral corresponding to
$\gamma$; $ T_{q_{\gamma}, m_{\gamma}} $ is the Taylor operator
expanding the integrand in small masses $\{ m_{\gamma} \}$ and
external momenta $\{ q_{\gamma} \}$ of the subgraph $\gamma$ 
; $ \circ$ stands for the convolution of the subgraph expansion
with the integrand $F_{G/{\gamma}}$. The sum goes over all
subgraphs $\gamma$ which (a) contain all lines with large masses, and
(b) are one-particle irreducible w.r.t. light lines.

  Individual integrals are specified by
the powers of the scalar denominators, called indices of the lines.
From the point of view of the asymptotic expansion method the
topology of the diagram is essential. All diagrams of the toy model
that contribute to the two-loop AMM can be classified in terms of $9$
prototypes (we omit the pure QED diagrams). These prototypes and
their corresponding subgraphs  involved in the asymptotic
expansion, are given in Fig.~\ref{fig33}. Almost all integrals occuring 
in the asymptotic expansion of the muon AMM in the SM can be evaluated 
analytically using the package SHELL2 \cite{SHELL2}. 

\subsection{The piloting C-program TLAMM}

For the automatic calculation  we have created a special
piloting program written in C.  This program
\begin{enumerate}
\item
reads QGRAF output;
\item
creates a file containing  the complete FORM program for calculating
each diagram;
\item
executes FORM;
\item
reads FORM output, picks out the result of the
calculation, and builds the total sum of all diagrams in a single
file which can be processed by FORM.
\end{enumerate}
Identifiers for vertices and propagators and the explicit Feynman
rules are read from separate files and then inserted into the FORM
program.  Because the number of identifiers needed for the
calculation of all diagrams at once may exceed the FORM capacity,
the piloting program retains for each diagram only those
involved in its calculation.

All initial settings are defined in a configuration file. The latter
contains information about the file names, identifiers of topologies,
the distribution of momenta, and the description of the model in
terms of the notation that is some extension of QGRAF's.
The program carries out the complete syntax check of all input files
except the QGRAF output.

There exist several options which allow one to process only the diagrams
\begin{itemize}
\item explicitly listed by number;
\item of a given prototype;
\item of a specified topology.
\end{itemize}
\noindent
There are also some debugging options.

\begin{figure}[h]
\begin{verbatim}
. . .
*--#[ d2:
*
     1
    *vx(E2(2),e2(-1),H(1))
    *vx(E2(3),e2(4),A(-3))
    *vx(E2(-2),e2(5),H(6))
    *vx(E2(5),e2(3),H(1))
    *vx(E2(4),e2(2),H(6))
*
*--#] d2:
. . .
\end{verbatim}
\vskip -47mm
\rightline{\vbox{\epsfysize=45mm \epsfbox{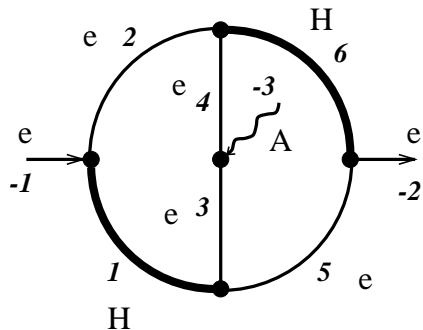}}}
\caption{\label{figd2} Diagram ``number 2''. {\tt e2} stand for fermion, 
{\tt E2} for anti-fermion, {\tt H} is a scalar, and {\tt A} the photon. 
External legs are numbered by negative numbers and vx stands for the vertex
for the particles in the argument list. }
\end{figure}

QGRAF generates the diagrams in symbolic form in terms of vertices.  
Fig.~\ref{figd2} shows one of the diagrams in the toy model with the 
corresponding QGRAF output.

The Feynman integrand generated by the C-program for this diagram reads:
\begin{eqnarray}
&&
\nonumber
\verb|g VF(mu)=|\\
\label{FI}
&&
\verb|E2_H_e2*F_F(-k2+p,q/2,me2)*E2_H_e2*F_F(k1-k2,q/2,me2)*|\\
\nonumber
&&
\verb|A_E_e(mu)*F_F(k1-k2,-q/2,me2)*E2_H_e2*F_F(k1,-q/2,me2)*|\\
\nonumber
&&
\verb|E2_H_e2*H_H(k1-p,0)*H_H(k2,0);|
\end{eqnarray}

A demo diskette for evaluating the toy model can be provided on request. 

\subsection{Project DIANA}

At present we have finished the main part of the more general program 
DIANA (DIagram ANAlyser). For generating Feynman diagrams we use 
again QGRAF. The program DIANA  consists of two parts:
\begin{itemize}
\item Analyzer of diagrams.
\item Interpreter of a special text manipulating language (TM);
\end{itemize}
The TM language is a very simple TeX-like language for creating
source code and organizing the interactive dialog.

The analyzer reads QGRAF output and passes necessary
information to the interpreter. For each diagram the interpreter performs
the TM-program, producing input for further evaluation of the diagram like 
for example (\ref{FI}).
Thus the program:

Reads QGRAF output and for each diagram it:
\begin{enumerate}
\item Defines the topology.
\item Looking for it in the table of
all known topologies and distributes momenta according to the current topology.
\item Creates a special internal representation of the diagram 
corresponding to the Feynman integrand.
\item Invokes the interpreter to execute the TM-program and passes to it
the necessary data.
\end{enumerate}

Executing the TM-program provides the possibility  to
calculate each
diagram using FORM or another formulae manipulating language, to do some numerical
calculation by means of FORTRAN, to create
a postscript file for the  picture of the current diagram, etc.

If we do not yet know all needed topologies,
we may use the program
to determine missing topologies that occur in the process.

The main goal of the TM-language is the creation of text files. 
In principle we could have used one of the existing languages, 
but we want a
very specific language: it should be powerful enough to create 
arbitrary program texts. On the other hand, it should be 
very simple and easy-in-use, so that even non-programmers can use it.

Similar to the TeX language, all lines without special escape - 
characters (``$\backslash$'') are simply typed to the output file. 
So, to type
``Hello, world!'' in the file ``hello'' we may write down the following
program:

\begin{verbatim}
\program
\setout(hello)
Hello, world!
\end{verbatim}

Each word
whose first character is an escape character will be considered as a command.
This feature makes this language very easy-to-use.

At present, we have finished the C-part of this project.

\section{Evaluation of scalar diagrams}

   In this second part methods for the evaluation of scalar three point functions
will be discussed. We will concern ourselves with the Taylor expansion
with respect to an external momentum squared and the large mass
expansion explained above. The efficiency of both approaches will be compared.

\subsection{Expansion of three-point functions in terms of
external momenta squared}

   Taylor series expansions in terms of one external momentum
squared, $q^2$ say,
were considered in \cite{Recur}, Pad\'{e} approximants
were introduced in \cite{bft} and in Ref. \cite{ft} it was demonstrated
that this approach can be used to calculate Feynman diagrams on their
cut by analytic continuation.
The Taylor coefficients are expressed in terms of ``bubble diagrams'', 
i.e.~diagrams with external momenta equal zero, which makes their
evaluation relatively easy.
In the case under consideration
we have two independent external momenta in $d=4-2 \varepsilon$ dimensions.
The general expansion of (any loop) scalar 3-point function with its
momentum space representation $C(p_1, p_2)$ can be written as
\begin{equation}
\label{eq:exptri}
C(p_1, p_2) = \sum^\infty_{l,m,n=0} a_{lmn} (p^2_1)^l (p^2_2)^m
(p_1 p_2)^n 
\label{2.2}
\end{equation}
where the coefficients  $a_{lmn}$ are to be determined from the given diagram.

For many applications it suffices to confine to the case
$p^2_1 = p^2_2 = 0$, which is e.g. physically realized in the case of the
Higgs decay into two photons ($H \to \gamma \gamma$) with $p_1$ and $p_2$
the momenta of the photons. In this case only the coefficients $a_{00n}$ are
needed. 

\begin{figure}[h]
\centerline{\vbox{\epsfysize=45mm \epsfbox{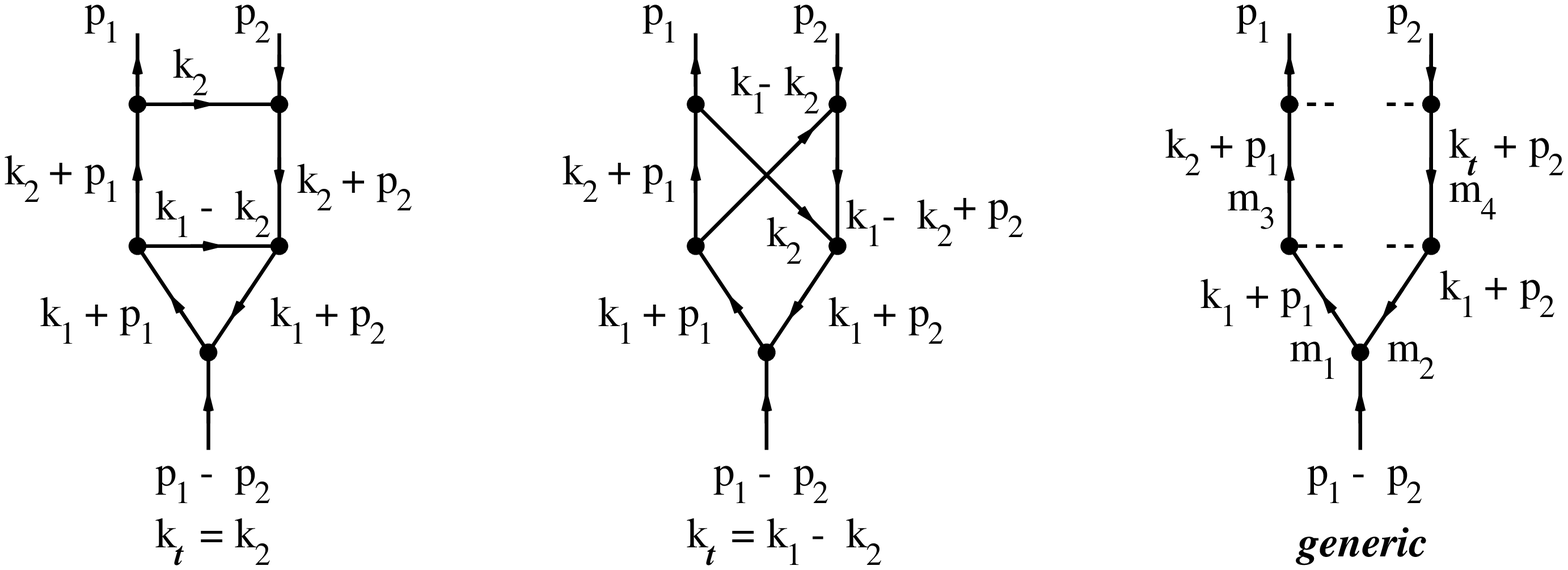}}}
\caption{\label{fig3} Planar and non-planar scalar vertex diagrams and 
their kinematics}
\end{figure}

  In the two-loop case we consider the scalar
integral ($k_3 = k_1 - k_2$, see also Fig.~\ref{fig3})
\begin{eqnarray}
\label{treug2}
\begin{array}{l}
C(m_1, \cdots, m_6; p_1, p_2) =\\
\\
\frac{1}{(i\pi^2)^2} \int
\frac{d^4 k_1 d^4 k_2}{((k_1 + p_1)^2 -m^2_1)((k_1 + p_2)^2 - m^2_2)
((k_2 + p_1)^2 - m^2_3) ((k_t + p_2)^2 - m^2_4) (k^2_2 - m^2_5)
(k^2_3 - m^2_6)}.
\end{array}
\label{2.4}
\end{eqnarray}
$k_t$ in line 4 (with mass $m_4$) depends on the topology: for the
{\it planar} diagram we have $k_t=k_2$ while for the {\it non-planar}
we have $k_t=k_3$. 



With obvious abbreviations for the scalar propagators: $c_i$ the $i^{th}$
scalar propagator of (\ref{treug2}) with $p_1=p_2=0$, we can quite generally
write for the $n^{th}$ Taylor coefficient \cite{ZiF}:
\begin{equation}
\label{e24}
(i\pi^2)^2 a_{00n} = \frac{2^n}{n+1} \int d^4 k_1 d^4 k_2 F_n \cdot
\frac{1}{c_1~ c_2~ c_3 ~ c_4~ c_5 ~ c_6}.
\end{equation}
The numerator $F_n$ can be written as 
\begin{equation}
F_n  = \sum^n_{\nu=0} c_1^{-(n-\nu)} c_3^{-\nu} \sum^n_{\nu^\prime=0}
c_2^{-(n-\nu^\prime)} c_4^{-\nu^\prime} \cdot A^n_{\nu\nu^\prime}
(k_1, k_2, k_t)\ .      \label{4.2}
\end{equation}
with
\begin{eqnarray}
\label{fullA}
\lefteqn{A^n_{\nu\nu^\prime} (k_1, k_2, k_t) = \sum^{\left[
\frac{\nu + \nu^\prime}{2}\right]}_{\mu =0}
\sum^{\left[ \frac{\nu}{2}\right]}_{\sigma =0}
\sum^{\left[ \frac{\nu^\prime}{2}\right]}_{\tau =0}
b^{n\mu,\sigma\tau}_{\nu\nu^\prime}}\\
& & \cdot (k^2_1)^{n-(\nu + \nu^\prime)+\mu} (k^2_2)^\sigma (k_1
k_2)^{\nu - \mu - \sigma + \tau}
(k_1 k_t)^{\nu^\prime-\mu+\sigma-\tau}
(k_2 k_t)^{\mu - \sigma - \tau} (k^2_t)^\tau , \nonumber
\end{eqnarray}
where the coefficients $b_{\nu\nu^\prime}^{n\mu,\sigma\tau}$ are given by
\begin{eqnarray}
b^{n\mu,\sigma\tau}_{\nu\nu^\prime}=
(n+1)\ 2^{\lambda - n - 1}\ (n - \nu)!\ (n-\nu^\prime)!
\nu !\ \nu^\prime       !
\frac{\Gamma (d-1)}{\Gamma (n+d-2)
\Gamma (n + \frac{d}{2})}  
&&\\
 \frac{2^{2\rho}}{\rho ! \sigma ! \tau !}
 \sum_{i=1}^{[ \frac{n}{2} ] + 1} (-4)^{i-1} (i-1)!\
\Gamma (n +
\frac{d}{2} - i) \nonumber 
&&\\
\nonumber \\
\sum_{j=\max (0,n_i-\nu)       }^{\min (n_i, n-\nu)       }  
  \sum_{k=\max (0,n_i-\nu^\prime)}^{\min (n_i, n-\nu^\prime)} 
         \frac{2^{j+k}}{(n_i - j - \rho)! (n_i - k - \rho)! (\rho + j + k-
n_i)!}\nonumber 
&&\\
\nonumber \\
\frac{1}{\sigma_j! (\nu_j - 2\sigma)! 
\tau_k ! (\nu^\prime_k - 2\tau)!} \ , \nonumber  
&&
\end{eqnarray}
with $\lambda=\nu+\nu^\prime-2\mu\,, 
\rho=\mu-\sigma-\tau\,,
\nu_j=\nu-n_i+j\,,\nu^\prime_k=\nu^\prime-n_i+k\,,
\sigma_j=\sigma - [j - (n -\nu) + (i-1)]$ and
$\tau_k=\tau - [k-(n-\nu^\prime) + (i-1)]$.

  For the {\it planar} diagram ($k_t=k_2$) (\ref{fullA}) simplifies 
and we have
\begin{equation}
\label{hernja}
A^n_{\nu\nu^\prime} (k_1, k_2) = \sum_{\mu = \max (0,\nu + \nu^\prime -n)}^{
\left[\frac{\nu + \nu^\prime}{2}\right]} 
a^{n\mu}_{\nu\nu^\prime} (k^2_1)^{n-(\nu + \nu^\prime)+\mu}
(k^2_2)^\mu (k_1~ k_2)^{\nu + \nu^\prime - 2\mu},
\end{equation}
with
\begin{equation}
a^{n\mu}_{\nu\nu^\prime} = \sum^{\left[ \frac{\nu}{2}\right]}_{\sigma =0}
\sum^{\left[ \frac{\nu^\prime}{2}\right]}_{\tau =0}
b^{n\mu,\sigma\tau}_{\nu\nu^\prime} \ , \label{ab}
\end{equation}

The coefficients $a^{n\mu}_{\nu\nu^\prime}$ are mass independent and have
been calculated with FORM up to order ${\varepsilon}^2$ ($d=4-2\varepsilon$)
and stored for the first 30 Taylor coefficients, i.e. they are given in terms
of rational numbers. For the {\it non-planar} case the situation is more
difficult since the storing of the coefficients 
$b^{n\mu,\sigma\tau}_{\nu\nu^\prime}$ with two more indices is practically
impossible for high Taylor coefficients. In many cases, however, if the 
threshold of the diagram under
consideration is high, only a few (say 8) Taylor coefficients are sufficient
for a high precision calculation of the diagram under consideration and
in such a case the direct calculation of the $b's$ in each case causes no
problem.

Finally all remaining integrals can be reduced to bubble-integrals of the type
\begin{equation}
V_{\alpha\beta\gamma}(\{m\}) = \int
\frac{d^dk_1 d^dk_2}{(k_1^2-m_1^2)^{\alpha}(k_2^2-m_2^2)^{\beta}
                                           ((k_1-k_2)^2-m_3^2)^{\gamma}} ,
\label{VBs}
\end{equation}
or to factorizing one-loop integrals.The genuine two-loop bubble integrals
are reduced by means of recurrence relations to
$V_{111}(\{m\})$. This can be done numerically for the
arbitrary mass case or also analytically for special cases like e.g.
only one non-zero mass. For details see \cite{recurrence},\cite{ZiF}. 
To perform the recursion numerically,
it is important to use the multiple precision FORTRAN by D.Bailey (\cite {Bail})
since tremendeous cancellations occur in this case.

We have in some detail presented one approach for the calculation of the
Taylor expansion of Feynman diagrams, others were worked out in
Refs. \cite{davt},\cite{Tara}. The latter one is particularly suited for
programming in terms of a formulae manipulating language like FORM.

\subsection{The method of analytic continuation}

Assume, the following Taylor expansion of a scalar diagram or a
particular amplitude is given 
$C(p_1, p_2,\dots)=\sum^\infty_{m=0} a_m y^m \equiv f(y)$
and the function on the r.h.s. has a cut for $y \ge y_0$.
In the above case of $H \to \gamma\gamma$ one introduces
$y =q^2/4m_t^2$ with $q^2=(p_1 - p_2)^2$ as
adequate variable with $y_0=1$.

 The method of evaluation
of the original series consists in a first step in a conformal mapping 
of the cut plane into the unit circle and secondly the reexpansion
of the function under consideration
into a power series w.r.t. the new conformal variable
\begin{equation}
\omega=\frac{1-\sqrt{1-y/y_0}}{1+\sqrt{1-y/y_0}}.
\label{omga}
\end{equation}

\begin{figure}[h]
\centerline{\vbox{\epsfysize=45mm \epsfbox{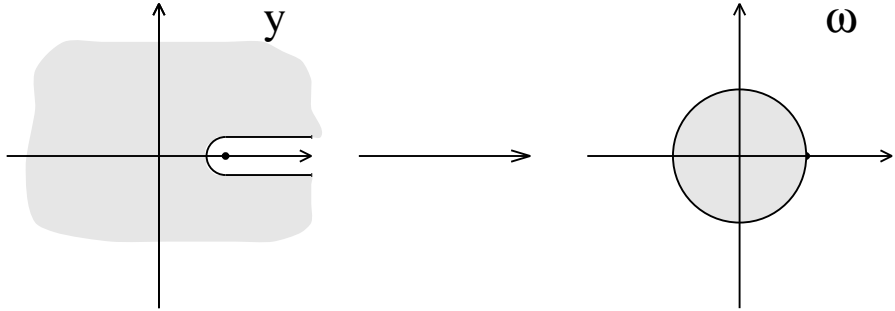}}}
\caption{\label{conf}Conformal mapping of the 
$y=q^2/4m_{\rm t}^2$-complex plane into the $\omega$-plane.}
\end{figure}

By this conformal transformation,
the $y$-plane, cut from $y_0$ to $+ \infty$, is mapped into the unit
circle (see Fig.\ref{conf}) and the cut itself is mapped on 
its boundary, the upper
semicircle corresponding to the upper side of the cut.
The origin goes into the point $\omega=0$.\\

  After conformal transformation it is suggestive to improve the
convergence of the new series w.r.t. $\omega$ by applying the 
Pad\'e method \cite{Sha},\cite{BGW}.
A convenient technique for the evaluation of Pad\'e approximations 
is the $\varepsilon$-algorithm of~\cite{Sha} which allows one
to evaluate the Pad\'e approximants recurrsively.

  Historically the first example considered was the 
two-loop three-point scalar ( {\it planar}) integral with the
kinematics of the decay $H \rightarrow \gamma \gamma$:
$m_6=0$ and
all other masses $m_i=m_t (i=1,..,5$). In this special case
all Taylor coefficients can be expressed in terms of
$\Gamma$ -functions. Later a closed expression for arbitrary
coefficients has been found in \cite{Tara}.

{\scriptsize
\hsize=11in\vsize=8in
\begin{table}[htb]
\begin{center}
\caption{Results on the cut $(q^2 > 4m^2_t)$
in comparison with~\protect\cite{DSZ}. }
\medskip
\def\.{&.}\def\pl{&$\pm$}
\halign{\strut\vrule~\hfil#&#~\vrule &~\hfil#&# &~\hfil#&#~\vrule  &~\hfil#&# &~\hfil#&#~\vrule
\cr
\noalign{\hrule}
$q^2$&$/m^2_t$ && [14/14] &&& & Ref.\protect\cite{DSZ}  && 
\cr
& && Re && Im && Re && Im \cr
\noalign{\hrule}  4&.01
& 11&.935 & 12&.699 & 11&.9347(1)
& 12&.69675(8) \cr
4&.10 & 2&.66245 & 9&.0955 & 2&.66246(2)
& 9&.0954(2)\cr
5&.0 & - 1&.985804823 &
2&.758626375& - 1&.98580(2) & 2&.758625(2)\cr
10&.0 &  -
0&.7569432708 & - 0&.0615483234 & - 0&.756943(1) & - 0&.061547(1)\cr
40&.0 & - 0&.045852780
& -     0&.0645672604 & - 0&.04585286(7) & - 0&.0645673(9)\cr
400&.0 &  + 0&.00008190
&       - 0&.0021670 & 0&.0000818974(3) & - 0&.002167005(3)\cr
\noalign{\hrule}}
\label{pade}
\end{center}
\end{table}}

Results for this kinematics on the cut are given in Table 1.
The process $H \to \gamma\gamma$ was investigated before in Ref. \cite{DSZ}.
For the master integral under consideration in \cite{DSZ}
all integrations but one could be
performed analytically and only the last one had to be done numerically.
Similarly, high precision is obtained on the cut in the approach of \cite{ft}.

Further examples of the efficiency of our method were given
in Refs. \cite{AIHENP} and \cite{Rheinsb}.
In the latter case a diagram of particular interest for the 
$Z \to b\bar{b}$ decay has been considered, namely a planar diagram with 
a low threshold: $m_1=m_2=m_5=m_6=m_b=4.5 GeV$ and $m_3=m_4=m_Z=91 GeV$.
If this diagram is to be evaluated on the Z mass-shell, 
a precision of only some four decimals can be obtained with 30 Taylor coefficients,
which is hardly good enough. On the other hand it is a very reasonable
approximation in this case to put $m_b=0$ from the very beginning
and thus consider diagrams with zero thresholds, which should be evaluated 
more easily with higher precision. These will be considered in the next
Sect.

\subsection{Two-loop vertex diagrams with zero thresholds}

\begin{figure}[h]
\centerline{\vbox{\epsfysize=45mm \epsfbox{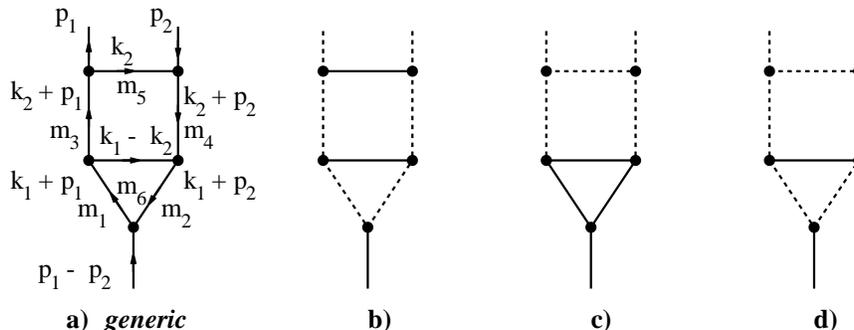}}}
\caption{\label{fig4} Planar diagrams with zero thresholds.
a)---generic;
b)---IR finite diagram; c),d)---IR divergent diagrams.}
\end{figure}

 Concerning the vertex diagrams, there are many different topologies
contributing to a 3-point function in the SM. For our
purpose of demonstra\-ting the method, we confine ourselves to the
planar case shown in Fig.~\ref{fig4}a.
Figs.~\ref{fig4}c,d presents infrared divergent diagrams.

As a typical (and very complicated case) we consider here Fig.~\ref{fig4}b
with equal non-zero masses \cite{FST}. There are two massless cuts
so that we shall have the double logarithm in the expansion.
The set of subgraphs in this case is given by
$\gm_1=\Gm$ and the higher terms with  $\gamma \ne \Gamma$ : $\gm_2=\{3456\},
\gm_3=\{1256\}$, and $\gm_4=\{56\}$ (in curly brackets: the line numbers of
the subgraphs $\gamma$ in (\ref{Lama}), line numbers are shown 
in Fig.~\ref{fig4}). 
Note that $\gm_3$ and $\gm_4$ are disconnected.
After summing up all four contributions we see that the double and
single poles in $\varepsilon$ cancel as well as the scale parameter $\mu$,
with the result $(x=q^2/M^2)$
\begin{eqnarray}
F_{\Gamma} (q^2,M^2) =
\frac{1}{M^4} \sum_{n=0}^{\infty} \sum_{j=0}^{2}
f_{jn} \ln^j(-x) x^n \nonumber \\
\label{AE22}
\equiv \frac{1}{M^4} \left\{ f_0(x) + f_1(x) \ln(-x)
+ f_2(x) \ln^2(-x)\right\} ,
\end{eqnarray}
where the $f_{jn}$ are now given in terms of rational numbers and $\zeta(2)$.
$f_2(x)$ can be summed analytically, yielding
\begin{equation}
f_2(x)= \bigl( \ln|1+x| -i\pi\theta(1+x) \bigr)^2/x^2  \,.
\end{equation}
Thus, we have to Pad\'{e} approximate $f_0$ and $f_1$ only.
Close to
the second threshold at $q^2=M^2$ the convergence is indeed excellent
(see Table \ref{padecut}) \cite{FST}. It should
be noted that for the physical application we have in mind, i.e. 
$Z \to b\bar{b}$,
this is just the case of interest. It is worthwhile to note the sharp increase
for low $q^2$ due to $\ln^2(-q^2/M^2)$.
\begin{center}
{\scriptsize
\begin{table}
\caption{Results for timelike $q^2$ for diagram \ref{fig4}b.}
\medskip
\def\.{&.}\def\pl{&$\pm$}
\halign{\strut\vrule~\hfil#&#~\vrule &~\hfil#&# &~\hfil#&#~\vrule  &~\hfil#&# &~\hfil#&#~\vrule
\cr
\noalign{\hrule}
$q^2$&$/M^2$ && [12/12] &&& & [15/15] && \cr
&& &Re && Im && Re && Im  \cr
\noalign{\hrule}
0&.05 & $+$2&.948516245  & $ $20&.938528 & $+$2&.948516245  & $ $20&.938528\cr
0&.1 & $-$1&.108116127  & $ $16&.04132127 & $-$1&.108116127  & $ $16&.04132127\cr
0&.5 & $-$4&.820692281  & $ $5&.066080015 & $-$4&.820692281  & $ $5&.066080015\cr
1&.0 & $-$3&.890154  & $ $1&.67549787 & $-$3&.890156  & $ $1&.67549788\cr
1&.5 & $-$2&.904588  & $ $0&.42979 & $-$2&.904581  & $ $0&.429778\cr
2&.0 & $-$2&.18294   & $-$0&.06976 & $-$2&.182981  & $-$0&.069728\cr
10&.0 & $-$0&.191   & $-$0&.208 & $-$0&.194   & $-$0&.215\cr
\noalign{\hrule}}
\label{padecut}
\end{table}}
\end{center}
Infrared divergent diagrams (Figs.~\ref{fig4}c,d)
have been successfully considered 
in \cite{Mainz}. For Fig.~\ref{fig4}c the 
convergence is indeed excellent again.
The different mass case, i.e. $m_6 \ne m_1=m_2$, had to be  calculated 
numerically according to the procedure outlined in Sect. {\it 2.1.}
For Fig.~\ref{fig4}d the convergence is by far not that good:
for $q^2/M^2 $ =~1 a precision  of 6 decimals is still obtained with a [15/15] 
approximant, while for $q^2/M^2 \simeq $ 10 only 2 decimals were obtained 
(see Table \ref{padecud}) where the ``numerical results'' were obtained
by the method of Ref. \cite{Kreimermethod}.
\begin{center}
{\scriptsize
\begin{table}
\caption{The finite part for diagram \ref{fig4}d.}
\medskip
\def\.{&.}\def\pl{&$\pm$}
\halign{\strut\vrule~\hfil#&#\vrule&~\hfil#&# &~\hfil#&#~\vrule  &~\hfil#&# &~\hfil#&#~\vrule
\cr
\noalign{\hrule}
$q^2$&$/m_6^2$ && [15/15] && && numerical results & &\cr
 $ $    & $ $         && Re      && Im && Re && Im   \cr
\noalign{\hrule}
0&.5 & $$81&.17501719  & $ $12&.06458720
& $ $81&.1750 & $ $12&.0644
\cr
1&.0 &  $  $ 17&.7659  & $ $ 19&.97799834 & $  $17&.7658 & $ $ 19&.9779 \cr
2&.0 &  $ -$  0&.6047  & $ $  7&.3759     & $ -$ 0&.604  & $ $  7&.376 \cr
10&.0 & $ -$  0&.572   & $ $  0&.023 & $ -$ 0&.570    & $ $ 0&.025 \cr
\noalign{\hrule}}
\label{padecud}
\end{table}
}
\end{center}
This much slower convergence implies that indeed it is very desirable
to have at least in the case of only one non-zero mass a compact
representation of the Taylor coefficients in terms of rational numbers,
i.e. not to have to go through the mashinery of calculating the coefficients
as described in Sect. {\it 2.1.} 
Such a possibility will be described in the next section.

\subsection{The Differential Equation Method}

  We saw that to obtain the expansion
of a diagram one has to go through a rather combersome machinery.
The more coefficients are asked for, the more efforts and machine
resourses are required. Thus it is very desirable to have analytic 
expressions for expansion coefficients whenever possible. 
This can be done
with the aid of the Differential Equation Method (DEM) \cite{DEMREW}
if only one non-zero mass occurs. 
The DEM allows one to get results for massive diagrams by reducing
the problem to diagrams with simpler structure. 

  Let us introduce a graphical notation for the scalar propagators
(in euclidean space-time) 

\vspace{-5pt}\hfill\\
\hspace*{3cm}
$\frac{\displaystyle 1}{\displaystyle (q^2)^\alpha}\,=\;\;$
\begin{picture}(30,10)(5,4)
\DashLine(0,5)(30,5){2}
\Vertex(0,5){1}
\Vertex(30,5){1}
\Text(15,7)[b]{$\scriptstyle\alpha$} 
\end{picture} 
$,\qquad \frac{\displaystyle 1}{\displaystyle (q^2+m^2)^\alpha}\,=\;\;$
\begin{picture}(30,10)(5,4)
\Line(0,5)(30,5)
\Vertex(0,5){1}
\Vertex(30,5){1}
\Text(15,7)[b]{$\scriptstyle\alpha$} 
\Text(15,3)[t]{$\scriptstyle m^2$} 
\end{picture} 
\vspace{-0pt}\hfill\\
\noindent
\vspace*{2mm}
$\alpha$ and $m$ refer to the index and mass of a line.
Then one can derive the following recurrence relation for
a massive triangle \cite{DEMREW} 

\vspace{-10pt} \hfill \\
\hspace*{1cm}
\trg{\alpha_1}{\alpha_2}{\alpha_3}
$(D-2\alpha_1-\alpha_2-\alpha_3) \,=\, -2 m_1^2\alpha_1\,\,\,$
\trg{\alpha_1+1}{\alpha_2}{\alpha_3}
\vspace{20pt} \hfill \\
\vspace{-0pt}
\vspace{-20pt}\hfill \\
\hspace*{1cm}
$+\alpha_2\Biggl(\;\;$
\trg{\alpha_1-1}{\alpha_2+1}{\alpha_3}
$-\;\;\;\;\;$
\trgleft{\alpha_1}{\alpha_2+1}{\alpha_3}
$\,-(m_1^2+m_2^2)\;\;\;\;$
\trg{\alpha_1}{\alpha_2+1}{\alpha_3}
$\Biggr) + (\alpha_2\leftrightarrow\alpha_3)$
\vspace{+5pt} \hfill \\
\noindent
along with some other graphical relations (see details in \cite{DEMREW}).

  Using this technique we analysed in \cite{DEM} the class of
3-point two-loop massive graphs. As an example for the diagram 
of Fig.\ref{fig4}c we get

\vspace{-30pt}\hfill\\
\hspace*{0cm}
$(D-4)J_{\rm 5c}\,=\, 2\!\!$
\begin{picture}(35,60)(5,27)
\DashLine(5,55)(35,55){2}
\DashLine(20,30)(5,55){2}
\DashLine(20,30)(35,55){2}
\BCirc(20,17.5){12.5}
\Line(20,5)(20,3)
\Line(5,55)(3,58)
\Line(35,55)(38,58)
\Text(35,15)[l]{$\scriptstyle 2$}
\end{picture}
$\!\! -\,2\;$
\begin{picture}(35,60)(5,27)
\DashLine(5,55)(35,55){2}
\DashLine(5,30)(5,55){2}
\Line(5,30)(35,55)
\Line(20,5)(5,30)
\Line(20,5)(35,55)
\Line(20,5)(20,3)
\Line(5,55)(3,58)
\Line(35,55)(38,58)
\Text(25,15)[l]{$\scriptstyle 2$}
\end{picture}
$ -\,4m^2$
\begin{picture}(35,60)(5,27)
\DashLine(5,55)(35,55){2}
\DashLine(5,30)(5,55){2}
\DashLine(35,30)(35,55){2}
\Line(5,30)(35,30)
\Line(20,5)(5,30)
\Line(20,5)(35,30)
\Line(20,5)(20,3)
\Line(5,55)(3,58)
\Line(35,55)(38,58)
\Text(29,15)[l]{$\scriptstyle 2$}
\end{picture}
$ -\,2m^2$
\begin{picture}(35,60)(5,27)
\DashLine(5,55)(35,55){2}
\DashLine(5,30)(5,55){2}
\DashLine(35,30)(35,55){2}
\Line(5,30)(35,30)
\Line(20,5)(5,30)
\Line(20,5)(35,30)
\Line(20,5)(20,3)
\Line(5,55)(3,58)
\Line(35,55)(38,58)
\Text(20,34)[l]{$\scriptstyle 2$}
\end{picture}
\hfill (16)\\ 
\vspace{15pt}

In the r.h.s. of (16) the last two terms can be combined, 
resulting in ${\rm d}J/{\rm d}m^2$ while for the second
term we proceed in turn as

\vspace{-30pt}\hfill\\
\hspace*{2cm}
$(D-4)$
\begin{picture}(35,60)(5,27)
\DashLine(5,55)(35,55){2}
\DashLine(5,30)(5,55){2}
\Line(5,30)(35,55)
\Line(20,5)(5,30)
\Line(20,5)(35,55)
\Line(20,5)(20,3)
\Line(5,55)(3,58)
\Line(35,55)(38,58)
\Text(25,15)[l]{$\scriptstyle 2$}
\end{picture}
$=$
\begin{picture}(35,60)(5,23)
\Line(20,5)(5,40)
\Line(20,5)(35,40)
\Curve{(5,40)(20,33)(35,40)}
\DashCurve{(5,40)(20,47)(35,40)}{2}
\Line(20,5)(20,3)
\Line(5,40)(3,43)
\Line(35,40)(38,43)
\Text(28,15)[l]{$\scriptstyle 2$}
\Text(20,49)[b]{$\scriptstyle 2$}
\end{picture}
$+$
\begin{picture}(35,60)(5,23)
\Line(20,5)(5,40)
\Line(20,5)(35,40)
\Curve{(5,40)(20,33)(35,40)}
\DashCurve{(5,40)(20,47)(35,40)}{2}
\Line(20,5)(20,3)
\Line(5,40)(3,43)
\Line(35,40)(38,43)
\Text(28,15)[l]{$\scriptstyle 2$}
\Text(20,31)[t]{$\scriptstyle 2$}
\end{picture}
$-$
\begin{picture}(35,60)(5,23)
\DashLine(20,5)(5,40){2}
\DashLine(5,40)(35,40){2}
\Curve{(20,5)(25,27)(30,36)(35,40)}
\Curve{(20,5)(25,9)(30,17)(35,40)}
\Line(20,5)(20,3)
\Line(5,40)(3,43)
\Line(35,40)(38,43)
\Text(22,25)[r]{$\scriptstyle 2$}
\Text(32,15)[l]{$\scriptstyle 2$}
\end{picture}
\hfill (17)\\
\vspace{25pt}

Thus we are left with simple diagrams (these can be done
completely by Feynman parameters) and the derivative
of the initial diagram w.r.t. $m^2$. The solution of the corresponding  
differential equation in terms of a series obtained from an
integral representation reads
\begin{eqnarray}
J_{\rm 5c}&=& -\frac{\Gamma^2(1+\e)}{{(q^2)}^2 {(m^2)}^{2\e}}
\sum^{\infty}_{n=1}  
\frac{(-x)^n \Gamma^2(n)}{\Gamma(2n+1)}
\Biggl[\frac{1}{\e^2} 
~-~ \frac{1}{\e} \biggl(\ln(x) +S_1(n-1)  \biggr) - \frac{3}{2}S_2(n-1)  
\nonumber \\ &-&   \frac{15}{2}S^2_1(n-1) 
 +4S_1(n-1)S_1(2n) - \zeta(2)
 - \ln(x)S_1(n-1) + \frac{1}{2}\ln^2(x)
  \Biggr],
 \nonumber
\end{eqnarray}\\
where
$$
 S_l(n)=\sum^{n}_{1}\frac{1}{k^l}
$$\\ 
 A similar formula was obtained for the 
diagram of Fig.\ref{fig4}d (see \cite{DEM}).

\subsection{Large mass expansion versus small momentum expansion}
  If there are two or more different masses involved the 
coefficients $a_{lmn}$ in (\ref{2.2}) are not just numbers 
any more but complicated functions of mass ratios. 
In this case one can try to perform a large mass (LM) expansion
rather than a small momentum expansion. 

\begin{figure}[h]
\centerline{\vbox{\epsfysize=28mm \epsfbox{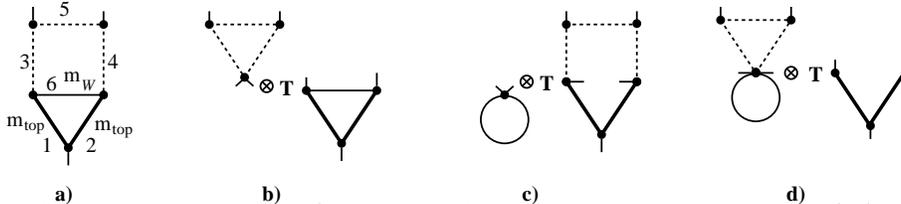}}}
\caption{\label{LMpic} A planar diagram a) with two different masses
         and subgraphs b),c) and d) contributing to the large mass
         expansion series. Dashed lines are massless; solid lines
         have light mass $m_W$; bold lines have heavy mass $m_{\rm top}$.
         $T$ stands for the Taylor expansion w.r.t. external momenta 
         and light mass.}
\end{figure}

  Here we just consider one example of a large mass expansion
for the 3-point function (a more detailed analysis will
be given in \cite{FKV}). Let us consider the diagram of Fig.~\ref{LMpic}a
with two non-zero masses $m_W$ and $m_{\rm top}$, contributing 
to $Z$ decay in the SM. Thus we have $q^2=m_Z^2$. 
One can use the procedure described in Sect. {\it 2.1.} to expand
this diagram in the external momentum squared. Then the actual parameter of
expansion is $q^2/m_{\rm top}^2$ and the radius of
convergence is given by the lowest (non-zero) threshold i.e.
$q^2=(m_{\rm top}+m_W)^2\ll m_Z^2$. Another possibility is to expand in the
ratios $q^2/m_{\rm top}^2$ and $m_W^2/m_{\rm top}^2$, though one will
see that this expansion does not work that well.

  The asymptotic expansion in the limit of the large 
mass is given by (\ref{Lama}). 
In our case there are 4 subgraphs that contribute,
i.e. $\gamma_1=\{123456\}$, $\gamma_2=\{126\}$
$\gamma_3=\{12345\}$ and $\gamma_4=\{12\}$ (see Fig.\ref{LMpic}a-d). 
By direct evaluation we find that there are induced
poles of the order $1/\epsilon^3$ in subgraphs while in the the sum 
they cancel which serves as a good check. For this particular
diagram the result of the LM expansion looks like
\begin{equation}\label{LM1}
   {\rm dia} =  \sum_{n=-1}^\infty A_n 
   = 
   \left( b^{(0)}_n
         +b^{(1)}_n L
         +b^{(2)}_n L^2
         +b^{(3)}_n L^3 \right)
   \left(\frac{1}{m_{\rm top}^2}\right)^n\, \nonumber
\end{equation}
with $L=\log(m_{\rm top}^2/\mu^2)$ and
$b^{(i)}_n$'s being known functions of $q^2,\,m_W^2$ and $\mu^2$.
  
  In Table \ref{LMtab} we give the results of the numerical analysis
for the graph at hand. We observed bad behavior of the
series (\ref{LM1}): namely 10 coefficients give only 3 stable figures.
Since the parameters of expansion $q^2/m_{\rm top}^2$
and $m_W^2/m_{\rm top}^2$ both are of order 0.25 one would 
expect better convergence. We have found that this bad convergence 
remains true
for some other graphs of the $Zb\bar b$ process \cite{FKV}. Moreover
no analytic properties are known in the $1/m^2$ variable
and thus no conformal mapping like (\ref{omga}) is now available to
improve the convergence. However we apply the Pad\'e summation.
Nevertheless the LM expansion for
3-point functions in such a regime looks very attractive 
since it can be rather easily implemented in formulae manipula\-ting 
languages like FORM even in the presence of more than one 
non-zero mass. In the worst case one achives 0.1\% accuracy
with 10 coefficiients. Our programs allow us to get 15-20
coefficients in a reasonable time.

{\scriptsize
\begin{table}[h]
\caption{Terms in the LM expansion corresponding to
formula (\ref{LM1}) at $\mu=m_{\rm top}=180\,, m_W=80\,, 
q^2=90^2$. The last two lines give  Pad\'e approximants obtained 
with 10 terms of the LM expansion and 8 Taylor coefficients 
of the small $q^2$ expansion.}

\begin{tabular}{|l|l|l|} \hline
$n$ & ${\rm Re}\,A_n$ & ${\rm Im}\,A_n$ \\ \hline
-1 & -6.35040 & + 29.98705\\
0 &  -3.85690 &  - 6.19256\\
1 &  -1.74681 &  - 4.34057\\
2 &   0.22329 &  - 1.30291\\
3 &   0.65892 &  - 0.21984\\
4 &   0.47058 &  + 0.00121\\
5 &   0.25376 &  + 0.01435\\
6 &   0.13099 &  + 0.00492\\
7 &   0.07311 &  + 0.00073\\
8 &   0.04517 &  - 0.00009\\
9 &   0.02984 &  - 0.00009\\
10 &  0.02042 &  - 0.00002\\ \hline
LM        & -9.996           & 17.9527  \\  \hline
small-$q$ & -9.9926682590236 & 17.95215366130182 \\ \hline
\end{tabular}
\label{LMtab}
\end{table}
}

\subsection{Conclusion}
  Any involved calculation in field theory nessesarily consists of
two parts: 1)automatic generation of Feynman diagrams and
source codes and 2)techniques of evaluating scalar Feynman diagrams.
In this paper we presented both. 
Still some efforts have to be applied to handle
the numerators of diagrams. We are working also on the automation
of the renormalization procedure.\\[0.5cm]

\noindent
{\noindent\bf Aknowledgements}
\vglue 0.2cm
  We would like to thank M.Kalmykov for helpful discussion.
M.T. and O.V. aknowledge the University of Bielefeld for
the warm hospitality.

\end{document}